\documentclass{llncs} 
\usepackage{graphicx}
\usepackage{amsfonts}
\usepackage{mathptmx}      % use Times fonts if available on your TeX system
\usepackage{hyperref} 
\usepackage{amsmath}
\usepackage{array}

\newcommand{\SEMCLA}{SemCla} % alg name semantic classifier 
\newcommand{\SEMCLABOLD}{\textbf{SemCla}} % alg name semantic classifier 
\newcommand{\SEMCAT}{SemCat} % alg name semantic categorizer
\newcommand{\SEMCATBOLD}{\textbf{SemCat}} % alg name semantic classifier 
\newcommand{\SEMCOM}{SemCom} % alg name commitee with semantic categorizer
\newcommand{\W}{$\mathfrak{W}$}

\usepackage{soul}

\begin{document}

\title{Semantic classifier approach to document classification}
 
   \author{Piotr Borkowski   \and
           Krzysztof Ciesielski \and
           Mieczys{\l}aw A. K{\l}opotek
   }
 
   \institute{%Piotr Borkowski, Krzysztof Ciesielski, Mieczys{\l}aw A. K{\l}opotek \at
                Institute of Computer Science, Polish Academy of  Sciences,\\ul. Jana Kazimierza 5, 01-238 Warszawa, Poland \\
                 Tel.: (+48) 22 380-05-00\\
                 Fax: (+48) 22 380-05-10\\
                 \email{piotrb, kciesiel, klopotek@ipipan.waw.pl}          
              }

\maketitle  
 \begin{abstract}
 
 In this paper we propose a new 
 document classification method, 
 bridging discrepancies (so-called semantic gap) between the training set and the 
 application sets of textual data. 
 We   demonstrate its superiority over classical 
 text classification approaches, 
 including traditional classifier ensembles. 
 The method consists in combining 
 a document categorization technique 
 with a single classifier 
 or a classifier ensemble (\SEMCOM{} algorithm - Committee with Semantic Categorizer).  
% %\keywords{semantic gap \and  semantic similarity \and  document categorisation \and  document classification \and classifier ensembles}
 \end{abstract}

\section{Introduction}
\emph{The text document classification} methods 
are well-established in the area of text mining.
Predominantly they have been derived from corresponding 
data mining techniques that were designed to handle long input data records.
Let us mention here for example Naive Bayes, Balanced Winnow and LLDA (to be described later). 
While these methods are quite successful in data mining
and were appreciated within text mining community,
one important drawback occurs related to the specific area of text mining.
While in data mining the meaning and the value range of individual attributes of an object are relatively well defined, in text mining it is not the case any more. 
Same content may be expressed in different ways, using different words (via synonyms, list of hyponyms) while the same word can express different things in different contexts. 
This would not be a big obstacle if not the fact that traditional techniques would require significantly larger bodies of training data, which makes an unbalanced sample much more likely. Not only because of the size of the data sample but also the heterogeneity of the data sources that need to be combined.  It is even worse when the trained classifiers need to be applied to unseen data which stems from a dataset that from the human point of view touches the same topic but from the computer point of view is written in a completely different style. 
This gives rise to so-called \emph{semantic gap}, that is though the training and application data sets are semantically similar, their syntactical and bag-of-words view differ. 
In such a case understanding the semantics of documents would be needed, which is unavailable for traditional data mining techniques.

In this paper we propose two new 
document classification methods, \SEMCLA{} (Semantic Classifier) and \SEMCOM{} (Committee with Semantic Categorizer),  
bridging the  semantic gap between the training set and the 
application sets of textual data. The methods consist in combining 
an unsupervised  \emph{document categorization technique} 
with a single classifier 
or a classifier ensemble. 
Via this component 
the traditional notion of document similarity
(based on angles between vectors in term space)
is amended to include the concept of
\emph{semantic similarity}. The notion of semantic similarity, as used in this paper, was described %introduced 
in \cite{SIIS:2011}.
Both methods introduced in the paper are based on our \SEMCAT{} (Semantic Categorizer) algorithm, that has also been introduced in \cite{SIIS:2011}.

In Section \ref{sec:related_work} we define the problem of document categorization and semantic classification and recall the work done on the subject by other researchers. In Section \ref{sec:categorization_method}  we describe our categorization methodology, \SEMCAT{}. 
Subsequently we show in Section \ref{sec:classical_classification},  how our categorization method can be used in various ways in the classical task of classification. 

In Section \ref{sec:experimental_setup} we explain 
the setup of experiments we performed to show the usefulness
of \SEMCLA{} algorithm in classification tasks. 
In subsequent Section \ref{sec:result} showing the results of these experiments, 
we  demonstrate   superiority of the semantic classification methods (\SEMCOM{} and \SEMCLA{}) over classical 
text classification approaches, 
including traditional classifier ensembles
for text classification tasks 
(Section \ref{sec:res_traditionalclass}) as well as 
in cases when the  so-called semantic gap occurs
(Section \ref{sec:res_semantic_gap}).
 
Section \ref{sec:conclusions} summarizes achieved results and outlines future research directions.
\subsection{Our contribution}
Our contribution in this paper is: 
\begin{itemize}
%[topsep=0pt,itemsep=-1ex,partopsep=1ex,parsep=1ex]
%[itemsep=10mm]
%\setlength{\itemsep}{-5pt}
%\setlength{\parskip}{0pt}
 \item constructing new supervised classifier based on unsupervised semantic document categorizator,
  \item demonstrating feasibility of the new classifier for bridging semantic gap between test and training set of data, 
  \item designing a heterogeneous committee that combines classical classifiers and the semantic classifier.
 %\item showing the possibility of training of the new classifier for a smaller data set.
\end{itemize}

\section{Previous work} \label{sec:related_work}

The task of \emph{categorization} 
is to  assign one or more labels (categories) to a document, or a group of documents (cluster labeling).
It finds multiple practical applications, especially for assisting in text retrieval task: in web page classification, e-mail and memo organization, expanding queries with new terms, expanding / improving ontologies, and many other.

The categorization task can be viewed formally as a special case 
of classification \cite{Sebastiani:2002,Sebastiani05textcategorization}, but with a couple of differences.
First of all, the number of categories significantly exceeds the number of classes in typical classification task. 
Categories may be flat and disjoint, but they may form a tree or even a hierarchy (acyclic graph). 
And more than one category may be assigned to a single document.
Therefore typical classification methods do not fit well to the task of categorization. 
Diverse other methods have been proposed to attack the problem of categorization. Some of them are based on clustering.  
The most popular representatives of this brand of approaches are   {Nonnegative Matrix Factorization (NMF)}, {Latent Semantic Analysis (LSA)}, {Probabilistic LSA (PLSA)}, and {Finite Mixture of Multidimensional Bernoulli Distributions}, described in  
\cite{SeppanenBM03}. 
Other researchers map the document contents to some semantic resources, in particular to Wikipedia (\W{}). This approach was exploited in  
\emph{WikipediaMiner Project}\footnote{\url{http://wikipedia-miner.sourceforge.net/}}, developed at the  University of Waikato in Hamilton, New Zeeland~\cite{Medelyan08,Milne08,MilneW08}. 
%http://www.cs.waikato.ac.nz/pubs/wp/2008/uow-cs-wp-2008-11.pdf
It uses \W{} topics as categories. 
Basic idea was key phrase indexing.
For terms from \W{}  their ``keyphraseness''~\cite{MihalceaC07} that is 
share of occurrences in \W{} links is computed.
Then these terms are searched in a document to be categorized.
Terms with multiple meanings are disambiguated (via some trained classifier) by choosing the meaning most close to the document topic.  
For training purposes documents annotated with such keyphrases have to be assigned categories. 
Then a classifier is trained.

 In  this paper we exploit our new unsupervised categorization method, \SEMCAT{}, introduced in \cite{SIIS:2011}. Contrary to WikipediaMiner, no classifiers are used, hence no training corpora need to be prepared. 
Also it is not based on \W{} links. Instead the category graph of \W{} is exploited. 
A novelty here is also the usage of more challenging Polish language \cite{sydow:debora-ismis12}.
Furthermore, we develop a classification method \SEMCLA{} suitable to apply for data  with semantic gap. 

The problem of ``semantic gap''  is understood in literature in many ways. We focus on the aspect encountered in text retrieval where data come for different domains. The next paragraphs give a brief overview of the approaches that have been proposed. 

 The article \cite{ramakrishna2012survey} shows a review of \emph{cross-domain text categorization} problem. Unlike the classical case, the training and the test data originates from different distributions or domains. This is very common in practical tasks because (especially for Polish language) we often do not have a suitable data set of labeled documents. Often what we have is a corpus which is topically related, but presents the same (or semantically similar) information in a different way, e.g. using different vocabulary. 
Many  algorithms have been developed or adapted for cross-domain text classification, there are conventional algorithms: Rocchio's Algorithm, Decision Trees like: CART, ID3, C4.5; Naive Bayes classifier, KNN, Support Vector Machines; and some novel cross-domain classification algorithms: Expectation-Maximization Algorithm, Probabilistic Latent Semantic Analysis (PLSA), Latent Dirichlet Allocation(LDA), CFC Algorithm, Co-cluster based Classification Algorithm \cite{wang2008using}. 
 
 Paper \cite{nguyen2010bridging} gives a general overview of the problem of semantic gap in information retrieval. Authors focus on two separate task: text and multimedia mining/image retrieval. Semantic gap in text retrieval is defined as a usage of different words (synonyms, hypernyms, hyponyms) to describe the same object. In the part about text retrieval authors concentrate on reorganizing search results by using post-retrieval clustering system. They work on search results (``snippets'') and enhance them by adding so called \emph{topics}. Topic is a set of words (they have similar meaning) that was as outcome of Probabilistic-Latent Semantic analysis or Latent Dirichlet Allocation on some external data collection. After adding a topic to the snippet they carry out clustering or labeling.
 
 In the paper \cite{ContentCategorizationRafi2012} authors propose a way to improve categorization by adding semantic knowledge from Wikitology (knowledge repository based on Wikipedia). They used various text representation and text enrichment techniques and used Support Vector Machine-SVM to learn a model of classification.

\section{Our taxonomy-based semantic categorization method} \label{sec:categorization_method}
Our taxonomy-based categorization method \SEMCAT{} was described in detail in \cite{SIIS:2011}, so below we present only its brief description.

\subsection{Outline of the algorithm}
\label{sec:outline_the_algorithm}
Suppose we have a taxonomy of categories (a directed acyclic graph with one root category) like Wikipedia ($\mathfrak{W}$) category graph or \textit{Medical Subject Headings} (MeSH) ontology\footnote{\url{https://www.nlm.nih.gov/mesh/}}.
We assume there is a set of concepts connected with the taxonomy, in the following way: every concept is linked to one or more categories. Every category and concept is tagged with a string label. Strings connected with categories are used as an outcome presented to a user. And those attached to concepts are used for mapping a text of document into the set of concepts.

For the experimental design we used $\mathfrak{W}$ category graph with the concept set of $\mathfrak{W}$ pages. Tags for $\mathfrak{W}$ categories were their original string names. Set of string tags connected with a single $\mathfrak{W}$ page consists of: lemmatized page name and all names of disambiguation pages that link to that page.  

In the  process of document categorization we remove stop words and very rare / frequent words, lemmatize, find  phrases and calculate
normalized $tfidf$ weights for terms and phrases. Calculation of a standard {\it term frequency inverse document frequency} is based on word frequencies from the collection of all $\mathfrak{W}$ pages. 
% Nastêpnie dla poszczególnych dokumentów wagi termów i fraz s¹ osobno normalizowane. W ten sposób promuje siê frazy, które mimo ¿e w tekœcie wystêpuj¹ rzadziej, nios¹ bardziej specyficzn¹ informacjê semantyczn¹ ni¿ pojedyncze s³owa.\\

Then we map document's terms and phrases into a set of concepts. In the case of homonyms, we disambiguate the concept assignment: we select the concept that is the nearest by similarity measure defined by Equations (\ref{eq:simLin}) and (\ref{eq:simPirro}) (see Section \ref{sec:semanticSimilarity}) to the set of concepts that was mapped in an unambiguous way. We investigated other methods of disambiguating e.g taking all meanings of ambiguous terms and weigh them accordingly. The results for various disambiguation methods are described in Section \ref{sec:methodology}. 

When every term in the document is assigned to a proper concept ($\mathfrak{W}$ page), then all concepts
are mapped to $\mathfrak{W}$ categories. In this way usually one term maps to more than one category, so we transfer the weight associated to that term proportionally to all its categories. Sum of weights assigned to the categories equals to sum of $tfidf$ for terms.  
The outcome of that procedure is a ranked list of categories with weights. 
In the last step we can transform the weighted ranking and / or 
choose top-$N$ categories out of it.\\

\subsection{Similarity measures} \label{sec:semanticSimilarity}
We use semantic measures for matching concepts ($\mathfrak{W}$ pages) and objects of the taxonomy ($\mathfrak{W}$ categories). We were inspired by the paper \cite{SemanticSimilarityMetric}. The semantic measures are based on: the unary function \emph{IC} (\emph{Information Content}) and binary function \emph{MSCA} (\emph{Most Specific Common Abstraction}). Their inputs are categories from a taxonomy.\\ 
Though superficially similar, our IC definition differs essentially from that proposed for WordNet. WordNet computes the IC for concepts based on the number of subordinated concepts. We compute the IC for categories, based on the count of concepts that belong to subordinated categories. So the IC of a category is weighted by the frequency of its usage in the language rather than by its definitional complexity.

%
% \Bem{
% To accommodate these measures to the particular form of $\mathfrak{W}$, a slight modification, taking into account the number of pages belonging to a category, is proposed here. We define $IC$ as follows\footnote{Pirro and Seco proposed formally identical formula, but they considered $s_k$ as the number of hyponyms of a given synset in Wordnet, and $N$ as the total number of synsets.}:
% }

For a given category $k$ we define $IC(k) = 1 - log \left( 1+s_k \right) / log \left( 1+N \right) $, where $s_k$ is the number of taxonomy concepts  in the category $k$ and all its subcategories, and $N$ is the total number of taxonomy concepts.
% So the highest values of $IC$ are assigned to the categories without any subcategory with only few concepts belonging to. The main category has the value of $IC=1-log(1+N)/log(1+N)=0$.
The main category has the lowest value of $IC=0$.

% For two given categories $k_1$ and $k_2$, the
% ${MSCA}(k_1,k_2) = \max\{IC(k): k \in CA(k_1,k_2)\}$,
% where $CA(k_1,k_2)$ is the set of super-categories for both categories $k_1$ and $k_2$. The properties of $IC(\dot)$ measure ensure that the category chosen is most specific amongst the common super-categories. 
% We define $MSCA(k_1,k_2)$ as the value of $IC(k^{*})$ for the category $k^{*}\in CA(k_1,k_2)$ maximizing $IC$.
% ${MSCA}(k_1,k_2) = \max\{IC(k): k \in CA(k_1,k_2)\}$,

For two given categories $k_1$ and $k_2$ we define $MSCA(k_1,k_2)$ as the category $k^{*}\in CA(k_1,k_2)$ (the set of super-categories for both categories $k_1$ and $k_2$) that maximize a value of the function $\{IC(k): k \in CA(k_1,k_2)\}$. The properties of $IC(\dot)$ measure ensure that the category chosen is most specific amongst the common super-categories. 

%We will denote by $k^{*}\in CA(k_1,k_2)$ category maximizing $IC$ value as $MSCA(k_1,k_2)$.

In the literature dealing with Wordnet many measures based on $IC$ and $MSCA$ have been proposed~\cite{SemanticSimilarityMetric}, including LIN and PIRRO-SECO similarity:

\begin{equation} \label{eq:simLin}
%\small
sim_\mathrm{Lin}(k_1,k_2) = \frac{2 \cdot MSCA(k_1,k_2)}{IC(k_1) + IC(k_2)}
\end{equation}

\begin{equation} \label{eq:simPirro}
%\small
%\begin{split}
sim_\mathrm{PirroSeco}(k_1,k_2) = \frac{1}{3} \Big(  3 \cdot MSCA(k_1,k_2)  - IC(k_1) - IC(k_2) +2 \Big)
%\end{split}
\end{equation}

Though analogous measures were defined for WordNet, our category similarity measures differ from those for WordNet because we defined IC and MSCA differently than in Wordnet.  Our definition is based on Wikipedia structure, hence we do not need to refer to WordNet.\\
We used the above measure for categories to define a similarity measures for concepts ($\mathfrak{W}$ pages). 
Similarity between pages $p_i$ and $p_j$ is computed by 
aggregation of the similarity between each pair of categories $(k_i,k_j)$  such that $p_i$ belongs to the category $k_i$ and $p_j$ to $k_j$:
% {\small
% $${sim_\mathrm{PAGE}(p_i, p_j) = max \{ sim_\mathrm{CAT}(k_i,k_j) : p_i \in k_i \wedge p_j \in k_j \}} $$
% %OPCJA BEZ PAGE $${sim(p_i, p_j) = max \{ sim_\mathrm{CAT}(k_i,k_j) : p_i \in k_i \wedge p_j \in k_j \}} $$
% }
\begin{equation} 
%\begin{split}
sim_\mathrm{PAGE} (p_i, p_j) = max \{ sim_\mathrm{CAT}(k_i,k_j) : p_i \in k_i \wedge p_j \in k_j \}
%\end{split}
\end{equation}

% We intend to look at this phenomenon more deeply in the future.

% 
% based on the
% 
% 
% So when computing similarity of \W{} pages $p_1,p_2$, we compute the similarity between each pair of categories $(k_1,k_2)$  such that $p_1$ belongs to the category $k_1$ and $p_2$ to $k_2$, and then we aggregate the results.
% 
%  of all pairwise category similarity of $p1$ and $p2$.

\section{Application to classification task} \label{sec:classical_classification}

%Poni¿ej znajduje siê krótki przegl¹d metod u¿ytych do nadawania kategorii dokumentom.

%The following chapter contains a brief overview of the %methods used to assign a set of categories to a document in a %supervised manner. 
%We call this process document labelling.  

%Kategoryzowano dokumenty tekstowe ze zbioru powsta³ego na bazie kolekcji DMOZ, dok³adny opis konstrukcji tego zbioru znajduje siê  w \ref{sec-zbior_dmoz}.
%Kategoryzowano dokumenty bêd¹ce podzbiorem kolekcji DMOZ, dokumenty te przypisane zosta³y do 15 kategorii. Kategorie te wybrane zosta³y w taki sposób, by istnia³y odpowiadaj¹ce im kategorie w ontologii kategorii Wikipedii. Dziêki temu do oceny wyników mo¿na stosowaæ miary wyliczane podstawie zale¿noœci miêdzy kategoriami w Wikipedii (patrz \ref{sec-miary_podobienstwa}).\\
% Dok³adna konstrukcja zbioru testowego opisana jest w rozdziale \ref{sec-zbior_dmoz}.

%Do nadawania etykiet ka¿demu z dokumentów u¿yto nastêpuj¹cych metod: stworzony przez nas algorytm do kategoryzacji, metody klasyfikacji takie jak Naive Bayes, %Balanced Winnow, Labeled LDA, komitet klasyfikatorów typu Bagging zbudowany z klasyfikatorów Naive Bayes i Balanced Winnow, niejednorodny komitet %klasyfikatorów (w sk³ad którego wchodzi algorytm do kategoryzacji).

In order to demonstrate the value of semantic categorization,
we exploited it as an ingredient (to a classifier ensemble)  in the classical classification algorithms and their committees, \SEMCOM{}, as well as an stand-alone classifier \SEMCLA{}.  

In this section we recall commonly known classification algorithms we used in our experiments.
%label a document we used the following methods: %our proprietary categorization method, supervised %classification methods such as 
These were Naive Bayes, Balanced Winnow, Labeled LDA, as well as the committees of classifiers (bagging type ensembles) built upon Naive Bayes classifier and Balanced Winnow. 
We describe also our own semantic categorization based classifier \SEMCLA{} and our heterogeneous committee \SEMCOM{} (containing both proprietary 
\SEMCAT{} 
%semantic categorization 
method and above-listed supervised classification methods).

%\subsection{Classification}

\subsection{Naive Bayes}\label{sec:bayes}
%Klasyfikacja metod¹ Naive Bayes (patrz \cite{Aas99textcategorisation}) pozwala na podstawie modelu, który budowany jest z u¿yciem wiedzy pochodz¹cej ze zbioru %danych treningowych, przypisaæ nowemu elementowi jedn¹ ze zdefiniowanych wczeœniej klas. W tym podejœciu ka¿dy dokument jest traktowany jako zbiór s³ów, w %którym nie uwzglêdnia siê kolejnoœci. Przyjmuje siê równie¿ upraszczaj¹ce za³o¿enie o niezale¿noœci wystêpowania s³ów w treœci dokumentu. Prawdopodobieñstwo %danej klasy $c$ przy ustalonym dokumencie $d$ obliczane nastêpuj¹co:

Naive Bayes classification method (cf. \cite{Aas99textcategorisation}) on the basis of knowledge derived from training data set, creates a probabilistic model assigning one of the predefined classes (i.e. labels) to a new observation (i.e document). In this approach, each document is treated as a bag of words, which does not take into account the order (syntax). Additionally, a simplifying assumption is made, that the individual words in the document are independent. The probability of a given class $c$ being assigned to a document $d$ is calculated as follows:

\[P(c|d)=\frac{P(c)\prod_{w \in d}P(w|c)^{n_{wd}} }{P(d)}, \]
%gdzie $n_{wd}$ jest liczb¹ wyst¹pieñ s³owa $w$ w dokumencie, $P(w|c)$ jest prawdopodobieñstwem wyst¹pienia s³owa $w$ w klasie $c$. $P(c)$ jest %prawdopodobieñstwem klasy $c$, które estymuje siê na podstawie frakcji dokumentów ucz¹cych nale¿¹cych do tej klasy. Wartoœæ $P(d)$ nie zale¿y od klasy, %dlatego przy klasyfikacji jest ono pomijane. Zaœ $P(w|c) = \frac{1+\sum_{d \in D_c} n_{wd}}{k+\sum_{w'}\sum_{d \in D_c} n_{w'd}, }$,  gdzie $D_c$ jest zbiorem %wszystkich dokumentów z klasy $c$, a liczba $k$ to rozmiar s³ownika.

\noindent
where $n_{wd}$ is the total number of occurrences of word $w$ in the document, a $P(w|c)$ is the probability of occurrence of a word $w$ in the class $c$. $P (c)$ is the probability of the class $c$ which is estimated based on the fraction of documents that belongs to this class. The value of $P(d)$ does not depend on the class, thus it is ignored for the purpose of document classification. Finally, $P(w|c) = \frac{1+\sum_{d \in D_c} n_{wd}}{k+\sum_{w'}\sum_{d \in D_c} n_{w'd}, }$, where $D_c$ is the set of all documents in the class $c$, and the number of $k$ is the size of the dictionary (i.e. the number of distinct words).

\subsection{Balanced Winnow} \label{sec:winnow}
%Opis algorytmu Balanced Winnow znaleŸæ mo¿na w pracach \cite{GroveBWinnow:2001} oraz \cite{LittlestoneBWinnow1988}. Istnieje kilka wersji tego klasyfikatora, którego idea opiera siê na algorytmie Perceptron (patrz \cite{Rosenblatt:1957}). Do eksperymentów wybrany zosta³ Balanced Winnow z racji jego obserwowanej wysokiej skutecznoœci. Dla ka¿dego s³owa algorytm przechowuje dwie wagi $w^+$ i $w^-$ na podstawie których oblicza siê przynale¿noœæ dokumentów do poszczególnych klas. Wagi dodatnie przemawiaj¹ na korzyœæ danej klasy, wagi ujemne przeciw niej. Ró¿nica miêdzy wagami ($w^+  - w^-$) jest ogóln¹ wag¹ zwi¹zan¹ z danym s³owem. Przyjmijmy, ¿e klasyfikowany dokument jest wektorem z wagami s³ów $x=(x_1 ,\ldots, x_n)$. Wtedy o przynale¿noœci dokumentu do danej klasy decyduje siê, je¿eli spe³niona jest nierównoœci: $\sum_{i=1}^n (w^+ - w^-)x_i > \theta$, dla ustalonego parametru $\theta$. Nauka klasyfikatora polega na modyfikacji wag, ale tylko w przypadku gdy dokument treningowy zosta³ Ÿle sklasyfikowany. Stosuje siê przy tym dwa parametry: promocji $\alpha>1$ oraz degradacji $0<\beta<1$. Je¿eli b³¹d polega na zaklasyfikowaniu dokumentu do klasy, do której on nie nale¿y (b³¹d na dokumencie negatywnym), wtedy nastêpuj¹co zmieniamy wagi s³ów bior¹cych udzia³ w klasyfikacji: $w^+:=\beta w^+, w^- := \alpha w^- $. Jeœli pope³nimy b³¹d na dokumencie pozytywnym (nie zaklasyfikujemy go do klasy, do której on nale¿y), wtedy dokonuje siê modyfikacji wag: $ w^+ := \alpha w^+, w^-:=\beta w^-$.

Balanced Winnow algorithm details can be found in \cite{GroveBWinnow:2001} and \cite{LittlestoneBWinnow1988}. Several versions of this classifier can be found in the literature. Main concept is based on the Perceptron algorithm (cf. \cite{Rosenblatt:1957}). For our purpose Balanced Winnow version of the algorithm was selected because of its high observed efficacy. For each word, algorithm stores two weights: $w^+$ and $w^-$, on the basis of which algorithm calculates document membership to each class (binary classification). Positive weights are in favor of a given class, negative weights against it. The difference between the weights ($w^+  - w^-$) is the overall weight associated with a given word. Assume that the classified document is a vector of words with the weights $x=(x_1 ,\ldots, x_n)$. Then the classification rule is based on the inequality $\sum_{i=1}^n (w^+ - w^-)x_i > \theta$, for a fixed value of the parameter $\theta$. Training of the classifier is based on weights modification, but only if the training document has been misclassified. Two parameters are introduced: promotion level $\alpha>1$ and degradation level $0<\beta<1$. If the error is to classify the document to the class to which it does not belong (negative document), then the weights of the words are modified as follows: $w^+:=\beta w^+, w^- := \alpha w^- $. If an error is made on a positive document (by not classifying it to the positive class), weight modification is as follows: $w^+ := \alpha w^+, w^-:=\beta w^-$.

\subsection{Labeled LDA}\label{sec:llda}

Labeled Latent Dirichlet Allocation (LLDA) is an extension of the popular -- among practitioners and theorists -- Latent Dirichlet Allocation model described in \cite{Blei:2003}. It is one of many probabilistic topic models useful in analyzing text documents. In particular the review of this subject can be found in \cite{steyvers2006probabilistic}.

LDA is an unsupervised method, where any document is treated as a probabilistic mixture of various topics. Resulting generative model is characterized by the discrete probability distribution of words within a given topic. The model assumes the following way to generate each document. The length $N$ of the document is selected (the Poisson distribution is used). Then the proportion of subjects making up the document is fixed (Dirichlet distribution randomizing the set of K topics). Subsequent words in the document are generated by the random selection of the topic (with a multinomial distribution generated above), and then within this topic (determining the distribution of words), a particular word is generated. Assuming such a method of generating each document in a given collection, LDA is trying to recreate a set of topics that are generating the observed collection. Labeled LDA method is a supervised variant, which relates every document label to a fixed subset of topics. LLDA algorithm is very similar to its unsupervised prototype, with the exception that the document topics are selected only from among those that correspond to the observed document labels -- details can be found in \cite{Ramage:2009}. There are other supervised variants of the LDA algorithm, such as Supervised LDA (\cite{blei2007supervised}). Authors selected LLDA in favor of Supervised LDA since in our experimental settings LLDA gave significantly better results. As part of future work it is planned to use also semi-supervised methods such as Partially Labeled Dirichlet Allocation (cf. \cite{Ramage:2011}).

% \cite{wang2009simulataneous}, , Rysunek \ref{fig:lda}
%
% \begin{figure}[h]
% \centering
% \includegraphics[trim = 0cm 0cm 2cm 0cm, clip,height=7cm]{fig/sLDA_sympleks.png}
% \caption{ Sympleks LDA}
% \label{fig:lda}
%
% \end{figure}

\subsection{Semantic classification}
\label{sec:semantic_classification}
  Below we present a description of a new \textit{semantic classifier} which we call \SEMCLA{}.
   It is based on a category representation of a document produced by \SEMCAT{} (see Section \ref{sec:outline_the_algorithm}), which is used in combination with semantic measures (see Section \ref{sec:semanticSimilarity}).   
   
   \subsubsection{Outline of the algorithm}
\begin{figure}
 
\centering

  \includegraphics[trim = 7cm 12.3cm 7cm 12.3cm, clip]{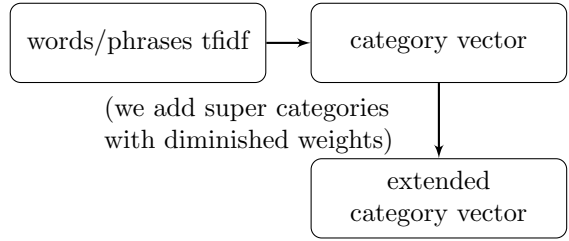}
  \caption{Single document category representation}
  \label{fig:document2category}
\end{figure}

      Recall that \SEMCAT{} uses words and phrases from the document to produce a list of categories with weights. This representation of a document can be considered as a vector of weights for all category from \W{} category structure. Therefore we call it \textit{vector of categories}. We use it to calculate cosine product. 
      We found out that the algorithm performs better when for each category from the vector of categories we add a super category of it (according to $\mathfrak{W}$ hierarchy) with weight equal the initial weight multiplied by a constant $\alpha$ (we used the value $\alpha=0.33$, we explain below how we calibrated this parameter). Thus we obtain the \textit{extended category vector}. This process is visualized in Figure~\ref{fig:document2category}.

\begin{figure}
\centering

\includegraphics[trim = 6.5cm 11.6cm 6.5cm 11.6cm,clip=true]{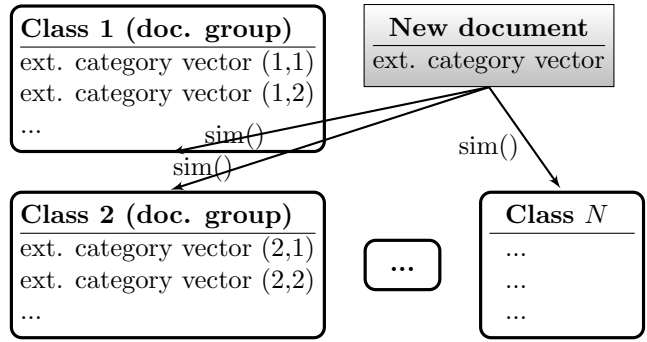} %width=\linewidth,
\caption{Categorization as a classification (\SEMCLA{} algorithm)}
\label{fig:semantic_classification}
\end{figure}

The semantic classification is made in the way described below and illustrated in Figure~\ref{fig:semantic_classification}.

  \begin{enumerate}%[itemsep=10mm]
      \item documents from training and test sets are categorized to obtain category vectors that represent their content,
    \item category vectors for all documents are changed into extended category vectors (for constant $\alpha$),
    %\item for each category from the vector we add a super category of it (according to $\mathfrak{W}$ hierarchy) with weight equal the initial weight multiplied by a constant $\alpha$ (we used the value $\alpha=0.33$, we explain below how we calibrated this parameter). We obtain the ``extended category vector''.
    %\item optionally we create a centroid for each document groups from training set
    \item we classify a new document (represented by its extended category vector) by finding the nearest group (in the sense of the cosine product) in the training set.
    \end{enumerate}

In the literature, a group to be compared with is represented by its centroid. Although the method with centroids works faster, it gives poorer results. Therefore results presented in Tables \ref{tab:varius_methods_dmoz_small_part1} -- %, \ref{tab:varius_methods_dmoz_small_part2},\ref{tab:semantic-gap-precision-CLASSICAL-METHODS}, 
\ref{tab:semantic-gap-precision-ENSAMBLES} are for \SEMCLA{} algorithms that find the nearest group using all documents from the group and taking average similarity.

\subsubsection{Finding optimal $\alpha$ parameter }
Value of optimal $\alpha$ was found in a separate experiment before the experiment discussed in the paper. It was conducted for \SEMCAT{} algorithm. We took 4 groups of documents from \url{kopalniawiedzy.pl}: \textit{astronomy-physics, psychology, medicine, technology} and drew at random $N=100$ documents from each of it. We did not use all document groups from this corpus, we chose 4 groups that were most different from each other. All documents were categorized with various values of $\alpha$ ranging from $0.0$ to $0.5$ (bigger $\alpha$ resulted in a significant deterioration of the outcomes). Then we calculated semantic similarity between categorized document (with different $\alpha$), sorted them and ranked. We chose the value of parameter $\alpha$ that maximizes difference between means of rank of documents from the same groups and those belonging to different groups. In other words, we found the value that separates best these groups of documents.

\subsection{Ensemble of classifiers} \label{sec:ensemble_of_classifiers}
%Stworzony zosta³ równie¿ algorytm klasyfikacji opieraj¹cy siê na komitecie klasyfikatorów. Dzia³a on w nastêpuj¹cy sposób. Dla ka¿dego dokumentu przeprowadza siê klasyfikacjê innym klasyfikatorem (mo¿e to byæ klasyfikator tego samego typu, ale zbudowany na innej próbie ucz¹cej). Otrzymane wyniki agreguje siê. W istniej¹cej implementacji mo¿na to robiæ na trzy sposoby: ka¿dy klasyfikator ma jeden g³os -- zliczane s¹ kategorie z najwiêksza liczb¹ g³osów; przy zliczaniu g³osów uwzglêdnia siê wagi z którymi zwracane s¹ wyniki klasyfikacji (ta opcja wymaga, ¿eby wszystkie stosowane klasyfikatory by³y tego samego typu); lub uwzglêdnia siê rangi elementów bêd¹cych wynikiem, zwracanym przez dany klasyfikator. W przypadku gdy dwie (lub wiêcej) kategorii otrzyma³o tyle samo g³osów, wynik wybierany jest losowo spoœród zwyciêskich kategorii. \\
%Do eksperymentów u¿yto rozmaitych wariantów komitetu klasyfikatorów, uczonych w opartu o ró¿ne podzbiory zbioru stron Wikipedii. Ze wszystkich stron nale¿¹cych do danej kategorii losowano 50 i na takiej podstawie stworzony by³ pojedynczy klasyfikator. U¿ywane w eksperymentach komitety w wersji podstawowej sk³ada³y siê z 20 klasyfikatorów, opartych na metodach Naive Bayes, Balanced Winnow (po 10 klasyfikatorów dla ka¿dej z metod) U¿yto metody agregacji, w której ka¿dy klasyfikator oddaje jeden g³os na jak¹œ kategoriê. Wiêcej informacji na temat komitetów klasyfikatorów naleŸæ mo¿na w ksi¹¿kach \cite{Elements:Hastie} czy \cite{Koronacki:StatSysUcz}.

The experimental setting was also based on the ensemble of classifiers. For each document the classification process is carried out
by every classifier in the ensemble (it may also be a classifier of the same type, but trained on a different learning sample). Then the results of all classifiers are aggregated as the final ensemble classifier. In the existing implementation this can be done in three ways: (a) each classifier has one vote -- category with the highest number of votes is selected; (b) votes counting additionally takes into account the weights of classification results (this option requires that all classifiers are of the same type); (c) ranks of the elements returned by the classifier are aggregated instead of raw votes or weights. In the case when two (or more) categories received exactly the same number of votes, the result is selected at random from among the winning categories. \\

\subsection{Heterogeneous committee of classifier with categorization method}\label{sec:heterog_committee}

%W zaproponowanym przez nas podejœciu, stworzono niejednorodny komitet klasyfikatorów w którym, oprócz metod Naive Bayes, Balanced Winnow oraz L-LDA, do komitetu do³¹czony zosta³ autorski algorytm kategoryzuj¹cy oparty na Wikipedii.\\
%Algorytm kategoryzuj¹cy nie mo¿e byæ ,,uczony'' na ró¿nych próbach (pos³uguje siê on danymi z ca³ej Wikipedii), dlatego do komitetu brano wynik z pojedynczego jego u¿ycia. Aby propozycje etykiet zwracane przez algorytm mia³y wiêkszy wp³yw na wynik klasyfikacji, jego g³osy uwzglêdnia siê z odpowiednimi wagami. Dodatkowo wykorzystano fakt, ¿e algorytm kategoryzuj¹cy jako wynik zwraca listê kategorii (a nie pojedyncz¹ kategoriê) i zbadano opcjê, w której do komitetu dodaje siê kilka kategorii z pocz¹tku tej listy, z odpowiednio malej¹cymi wagami.

In our new approach, we developed heterogeneous committee of classifiers \SEMCOM{} that contains the supervised methods of Naive Bayes, Balanced Winnow, LLDA and our proprietary unsupervised categorization method \SEMCAT{} utilizing taxonomy of \W{} categories. \\
Categorization method is unsupervised, and thus it cannot be trained on different samples in a similar manner to supervised classifiers (categorization method utilizes data from the complete \W{} taxonomy). For this reason the committee contained only one instance of the categorization algorithm. In order to increase the impact of \SEMCAT{} on the final results of the committee as a whole, categorization votes were counted with the higher weight. In addition, one should take into account that the categorization algorithm returns a ranking of categories (not only a single category). Thus, in the experimental settings we included a variant of the committee in which categorization method add more than one category with the highest rank in the list (and the correspondingly decreasing weights).

\subsection{Remarks on denotation of classifier and ensemble parameters and composition}

Experimental setting exploited several variants of the ensembles, trained on a different subsets from the training set 
(\W{} pages for Table \ref{tab:varius_methods_dmoz_small_part1}, \ref{tab:varius_methods_dmoz_small_part2} and groups of news for Tables \ref{tab:semantic-gap-precision-CLASSICAL-METHODS}, \ref{tab:semantic-gap-precision-ENSAMBLES}).

For the \textit{classical classification task} (Table \ref{tab:varius_methods_dmoz_small_part1} and \ref{tab:varius_methods_dmoz_small_part2}) from all \W{} pages belonging to a single category [$S = 50,100,200$] pages were drawn at random and on the basis of such a sample, a single classifier was trained. 
For a given set of classes, into which documents are to be classified, we choose \W{} categories that represent these classes.
We will call them \W{} class categories.
When we choose \W{} documents for training, we can choose either documents that have \W{} categories identical with the \W{} class categories or their sub-categories.
We say that we choose level~1 ($L=1$) documents, if for each document at least one of its categories is identical with class category.
If we choose $L=2$ documents,  then we choose additionally also documents  that have categories being direct subcategories of the class category.
%Parameter $L$ stands for the level of \W{} documents used to compose training sample. For example, $L=p$ represents all documents belonging to \W{} categories that is distant from \W{} root category by $p$ (in the sense of shortest path in \W{} category graph). 
Vector of numbers following the \SEMCAT{} represents weights attached to top-3 categories inserted into committee. We chose them among all categories produced by \SEMCAT{} algorithm (e.g \SEMCAT{}:$(7,5,3)$ means that we put top three categories from semantic categorizer with weights $7,5$ and $3$).\\
For the \textit{semantic gap task} we used $S=50$ for Table \ref{tab:semantic-gap-precision-CLASSICAL-METHODS} and $S=200$ for Table \ref{tab:semantic-gap-precision-ENSAMBLES}. Experimental committees consisted of 25 classifiers based on Naive Bayes and Balanced Winnow methods. Aggregation variant was the one in which each classifier is voting on one category only. More information on ensemble methods can be found in \cite{Elements:Hastie}.

\section{Experimental setup} \label{sec:experimental_setup}

\subsection{Performed experiments} \label{sec:performed_experiments}
We performed two types of experiments, their
results are reported in Tables 
\ref{tab:varius_methods_dmoz_small_part1} -- %, \ref{tab:varius_methods_dmoz_small_part2} and \ref{tab:semantic-gap-precision-CLASSICAL-METHODS},
\ref{tab:semantic-gap-precision-ENSAMBLES}.
The first experiment aimed at demonstrating 
that adding a semantic categorizer to 
a committee of traditional classifiers improves 
the classification correctness 
in \emph{classic classification task} 
(Table \ref{tab:varius_methods_dmoz_small_part1}, \ref{tab:varius_methods_dmoz_small_part2}).
The second experiment was designed to show that 
a semantic categorizer is capable of \emph{bridging 
semantic gap} between the training data and the test data
(Table \ref{tab:semantic-gap-precision-CLASSICAL-METHODS}, \ref{tab:semantic-gap-precision-ENSAMBLES}).

\subsection{Benchmark data sets} \label{sec:dmoz_dataset}

For experimental purpose we used two different benchmark data sets. We needed different datasets because of various nature of the investigated problems.\\ 

\noindent
\textbf{Benchmark used for classification comparison}.

The benchmark data set was based on Polish subdirectory of DMOZ taxonomy / Open Directory Project \url{http://www.dmoz.org}. It contains  1063 text files of  Polish web pages just with html tags removed. Selected documents belong to 15 directories that map into $\mathfrak{W}$ categories. They are: 
astronomy, biology, economics, philosophy, physics, graphics, history, linguistics, mathematics, education, politics, law, religious studies, sociology, technology. None of these categories is a subcategory of another one in the $\mathfrak{W}$ taxonomy. 
 We omitted a few cases of multi-labeled documents. For the benchmark documents the reader is referred to the benchmark web page\footnote{ \url{http://www.ipipan.waw.pl/~kciesiel/iis/DMOZ_PL_taxonomy.zip}}. The various options of categorization setting cause the number of categorized document differs. For calculating the results we choose a set of documents that was categorized by every algorithm.\\

\noindent
\textbf{Benchmark containing data with semantic gap.}

 The second benchmark was made of documents downloaded from various news page. It consists of training and evaluation part, they come from various domains. We used  separate collections to achieve different wordings in each of them. The training set consists of news from
 the popular science portal \url{kopalniawiedzy.pl} merged with documents from one directory from \url{forsal.pl}  -- the domain about finance and economy. Below we show more detailed description of the training set. 
 %Datasets used for for training:  directory: classifiers was $N=50$ documents form each. 

\begin{itemize}
  \item documents from  \url{kopalniawiedzy.pl}: \textbf{astronomy-physics} N=283; \textbf{medicine} N=2979; \textbf{life science} N=3122; \textbf{technology} N=4861; \textbf{psychology} N=1733; \textbf{humanities} N=244, 
   % \textbf{astronomia-fizyka, medycyna, nauki-przyrodnicze, technologia, psychologia, humanistyka}
\item  documents from \url{forsal.pl} from the directory \textbf{ Gie{\l}da (Stock exchange)} N=1987.

\end{itemize}

For evaluation we downloaded directories from \url{www.rynekzdrowia.pl} (containing medical news) and merged it with economical documents from \url{www.forsal.pl} and \url{www.bankier.pl} (market, finances, business). Datasets used for evaluation:
\begin{itemize}
  \item directories from \url{www.rynekzdrowia.pl}: \textbf{Ginekologia (Gynecology)} N=1034; \textbf{Kardiologia (Cardiology)} N=239; \textbf{Onkologia (Oncology)} N=1195,
  
  \item directories from \url{www.forsal.pl}: %\textbf{Gie{\l}da (Stock exchange)} N=1986; 
  \textbf{Waluty (Currencies)} N=2161; \textbf{Finanse (Finances)} N=1991, 
  \item documents from \textbf{www.bankier.pl} N=978.
  
\end{itemize}

\subsection{Efficiency measures}\label{sec:efficiency_measures}
To assess the efficiency of the studied algorithms we use two different measures. The first one is commonly used standard \emph{precision} measure, the second one is modified precision based on similarity measure Lin (Equation (\ref{eq:simLin}) in Section \ref{sec:semanticSimilarity}). The difference is in using Lin measure instead of indicator function. For documents $d_1, ... , d_n$ with real categories $categ(d_i)$ and its prediction $pred(d_i)$ the {\it Lin precision} is defined as: $\frac{1}{n}\sum_{i=1}^n Lin(categ(d_i),pred(d_i))$. The motivation for using the latter measure is that standard precision does not take into account the dependence between categories. In case when we make a wrong prediction we would like to know how much predicted category is different from the real one.

\subsection{Classical classification task}\label{sec:methodology}

The first part of experimental work concerned comparison of various methods of text classification. We proceeded on documents from DMOZ corpus with 
fixed set of labels described in Section \ref{sec:dmoz_dataset}. 
Documents were divided into separate groups based on their text length measured by the number of characters~($C$): {\it short} ($1000 \le C<2000$), {\it medium} ($2000 \le C<10,000$), {\it long} ($10,000 \le C$). Files shorter than $1000$ characters were not processed. Results for various classification methods are presented in Tables \ref{tab:varius_methods_dmoz_small_part1}, \ref{tab:varius_methods_dmoz_small_part2}. They were divided by a file size and efficiency measure. 
Methods based on categorization algorithm return a list of weighted \W{} categories. Therefore we transformed 
the outcome categories into the target set of 15 categories and took only one category with the highest weight.
Categorization was based on a selection of 10 words (only nouns) / phrases with highest $tfidf$ from the document.
The experiments were performed for different values of parameters, but other settings gave worse results.

In Table \ref{tab:varius_methods_dmoz_small_part1} first four rows present various modifications of categorization method. 
The difference between them is in the method of disambiguation of ambiguous page. The first row presents standard disambiguation method (see Section  \ref{sec:categorization_method}). The next two methods find a set of pages that map unambiguously. Then for every ambiguous page we find all of its mappings to potential meanings. Then we figure out their distances to the set and sort them into descending order. Subsequent possible meanings are given various weights depending on their rank $i$: ($1/2^i$, $1/i$  or uniformly). 
All of these options gave similar means, so we used paired t-test to compare them. As a reference we used basic disambiguation method. Methods with weighting $1/2^i$ and $1/i$ do not differ significantly. Method with uniform weights differs.

All of these methods took only nouns from the document. We developed two options of mapping words into titles of \W{} page. We remove (or not) from the set of possible page those of them that do not match in an exact way. The option ``exact matching'' worked slightly better (although not significantly), so we present it. Then we present individual classifiers followed by the ensembles of classifiers.
Subsequent results are for heterogeneous committee. 

\subsection{Classification for data with the semantic gap} \label{sec:semantic_gap}
The second experiment  focuses on the problem of semantic gap which is observed in classification of data from different domains. 
For such data often two documents express the same concepts, but as they use different wording (because of existing of synonyms, hypernyms, hyponyms), the conventional classification / clustering algorithms, based on standard bag-of-words approach, do not work well. Such classifiers often do not recognize different linguistic representations for test and training set. Some works relating to the problem were presented in Section \ref{sec:related_work}.
Our approach, thoroughly presented above, is different from them.\\
There are other linguistic phenomena such as ellipsis, paraphrase and other. We focus on synonyms, hypernyms, hyponyms because of Wikipedia structure on which our algorithm is based. We deal with hyper-/hyponyms relation because of \W{} category graph structure we operate on. This graph is built on these kinds of relations.\\
With synonym relation we cope during the phase of mapping words/phrases from the text into \W{} pages. The string set attached to a single \W{} page contains the page title and all it's synonyms. They are extracted from all names of disambiguation pages that point to this particular page.\\
For experimental design (see Table \ref{tab:semantic-gap-precision-CLASSICAL-METHODS}) we used standard classification methods in different settings. As an input for them we used: 1.{ \textit{terms}} -- terms from the document; 2. {\textit{categories}} -- categories for a given document produced by \SEMCAT{};  3. {\textit{concepts}} -- set of disambiguated concepts (\W{} page id) produced during \SEMCAT{} algorithm. \\ In Table \ref{tab:semantic-gap-precision-ENSAMBLES} we present \SEMCLA{}, ensembles and the heterogeneous committee with semantic classifier.

%Possible difficulties with classification for data with the semantic gap:
% \begin{itemize}
% \item semantic gap: usage of different words (synonyms, hypernyms, sets of hyponyms) to express the same concept in the training and test set of documents
% 
% \item document representation based on standard bag-of-words approach is problematic for data with the
% semantic gap.
% \item classifiers do not recognize different linguistic representations f
%or test and training set.
% \end{itemize}
% 
% Methods that use a semantic representation for documents, potentially applicable  in such cases:
% \begin{itemize}
% \item usage of semantic resources
% \item appropriate definition of similarity between concepts required
% \item major issue: the size of the conceptual space too big for the training sample available; need of appropriate concept projection onto a representative subspace
% \end{itemize}

\section{Results}\label{sec:result}

\subsection{Classical task} \label{sec:res_traditionalclass}

As can be seen in Tables \ref{tab:varius_methods_dmoz_small_part1}
the best method among the considered \SEMCAT{} algorithms 
is the one where upon mapping of terms/phrases to \W{} pages 
the ranking of pages corresponding to a term is computed
and all of them are taken into account using appropriate weights. 
The version using only unambiguous terms and phrases has the poorest performance. 
Modifications of the base method (variants of fitting, shifting the stage of category projection) do not lead to significant changes in performance.  
% Z zaproponowanych modyfikacji algorytmu kategoryzuj¹cego najlepiej dzia³a metoda, w której przy odwzorowaniu s³owo/fraza na strony Wikipedii wylicza siê ranking potencjalnych stron odpowiadaj¹cych termom, a nastêpnie uwzglêdnia siê wszystkie z nich z odpowiednimi wagami. Metoda korzystaj¹ca tylko ze s³ów/fraz jednoznacznych dzia³a najgorzej, zaœ ró¿ne modyfikacje metody podstawowej (ze wzglêdu na rodzaje dopasowania, zmiany momentu rzutowania na kategorie wynikowe) daj¹ zbli¿one do siebie wyniki.\\

Though \SEMCLA{} outperforms individual non-semantic classifiers,
one can see that a classical classifier ensemble is able to outperform 
\SEMCLA{}.  

Therefore we turned to considering the impact of inclusion of \SEMCAT{} into an ensemble of classical classifiers.   

% Najskuteczniejsz¹ z metod okaza³a siê metoda niejednorodnego komitetu klasyfikatorów. Z tego powodu oddzielnie zaprezentowano te¿ ró¿ne zestawienia takich komitetów.

The size of the ensemble (25x Balanced Winnow + 25x Bayes) guarantees the stability of the results under various selections of the random training samples.

% \item wielkosc komitetu 25x winnow + 25x bayes jest tak dobrana, ze losujac inny komitet o tych samych parametrach uzyskuje zblizone wyniki. Dlatego nie bralismy wielu ensambli tylko prezentujemy jeden.
Experimental settings included: various levels of \W{} category graph used to create training samples [Level=${1,2,\infty}$] as well as various sample sizes per category [$S = {50,100,200}$]. Optimal results (presented in Tables \ref{tab:varius_methods_dmoz_small_part1}, \ref{tab:varius_methods_dmoz_small_part2}) were achieved for Level=$2$ and $S=200$. In particular, Level=$\infty$ led to noticeably worse performance, since \W{} documents selected in the random sample were vaguely related to the desired topic (category). 

% \item  badalismy rozne opcje komitetu 25x winnow + 25x bayes: rozne poziomy dokumentow level=${1,2,\infty}$ i rozne liczebnosci probek dokumentow dla kazdej kategorii $N = {50,100,200}$. Najlepiej wypadalo ustawienie level=$2$ i $N=200$. \\
% Tworzac komitet ``ensemble'' biorac dokumenty z level=$\infty$ wyniki sie pogarszaja, bo dokumenty brane do nauki sa luzno zwiazane z zadanym tematem

On the other hand, in every investigated case, results for the Level=1 were worse than for Level=2, since the randomization of the sample for each instance of the 
 classifier was too low (the number of the \W{} documents on level 1 was not sufficient to make a sample).\\ % \item systematycznie wyniki dla komitetu Winnow+Bayes uczone na plikach lev=1 byly gorsze, dlatego prezentujemy tylko dla lev=2
\indent Ensemble of classical classifiers was extended with \SEMCAT{} (Table \ref{tab:varius_methods_dmoz_small_part2}) using various weights for 1st, 2nd \& 3rd category in the \SEMCAT{} ranking. This setting required further investigation, 
but usually weights 14/10/6 led to the best classification results. Higher weights caused worse results. Extended ensemble 25x Balanced Winnow + 25x Bayes + \SEMCAT{} with Level=$2$, $S=200$ and weights 14/10/6 usually was the optimal setting (with an exception for shortest documents).\\ 
% \item dodajac do roznych ensambli nasz klasyfikator testowalismy rozne wagi. Nie jest latwo ocenic ktory zestaw wag jest najlepszy jednoznacznie-- chyba najlepszy jest 14 10 6 (zaraz za nim 10,5 7.5 4.5). Powyzej tych wag obserwuje sie spadek jakosci. Najlepsze wyniki sa dla komitetu $Level 2$ $N=200$ + kategoryzator 14 10 6 -- ale to zalezy od miary i wielkosci pliku.
\indent Further extension of the ensemble with LLDA classifier did not improve the results, both in the case of base ensemble (25x Balanced Winnow + 25x Bayes) and the semantic ensemble 
that included \SEMCAT{} algorithm.\\ 
% \item dodawanie LLDA nie pomaga. W poprzednich wynikach (z artykulu 2 lata temu) LLDA - nieznacznie poprawialo wyniki. Teraz uzywamy optymalnie dobranego komitetu (25x Bayes + Winnow) i zapewne dlatego LLDA juz niewiele wnosi. (DO LLDA uzylem jednej z probek na ktorej trenowany byl Bayes do zestawu 25x Bayes).
% \item tak samo dodanie LLDA do komitetu heterogenicznego nie pomaga wiele
%\subsection{Discussion} \label{sec-dyskusja}
\indent Presented experiments lead to the following conclusions: the best results were achieved for ensembles that beside standard
classification methods (25xBayes + 25xBalanced Winnow) a semantic method was
included (either \SEMCAT{} or \SEMCLA{} algorithm). Surprisingly, adding
more varied set of standard classification methods (Naive Bayes,
Winnow and LLDA) did not improve quality of the ensemble.\\
\indent Ensemble of 25x\SEMCLA{} classifiers in most cases does not perform significantly better than a single \SEMCLA{}.
It is mainly due to low variance of the individual voting methods within the ensemble.

\begin{table*}[h]

%\large
%\small
%\footnotesize
%\scriptsize
%|p{4.3cm} -- koluma o stalej szerokosci w cm
\caption{Average values of various precision measures for \textit{DMOZ small} dataset. Parameter $L$ stands for a level of \W{} documents used for training sample, $S$ is a sample size per each group of documents. 25x(B,W) stands for an ensemble of 25 Bayes an 25 Balanced Winnow classifiers. Vector of numbers that follows the SemCat represents weights attached to top-3 categories inserted into committee.}
\begin{tabular}{|l|l|l@{}|l@{}|l@{}|l@{}|l@{}|l@{}|}

% \begin{tabular}{|@{}c@{}|@{}c@{}|@{\hspace{2pt}}c@{\hspace{2pt}}|@{\hspace{2pt}}c@{\hspace{2pt}}|@{\hspace{2pt}}c@{\hspace{2pt}}|@{\hspace{2pt}}c@{\hspace{2pt}}|@{\hspace{2pt}}c@{\hspace{2pt}}|@{\hspace{2pt}}c@{\hspace{2pt}}| @{}}

\hline

\multicolumn{2}{|@{}c@{}|}{} & \multicolumn{3}{c|}{Lin precision} & \multicolumn{3}{c|}{Precision} \\
\hline
Method & Description &\footnotesize{short}&\footnotesize{medium}&\footnotesize{long} &\footnotesize{short}&\footnotesize{medium}&\footnotesize{long} \\
 \hline
%Wiki podstawowa
% \\ \hline

SemCat &\SEMCATBOLD{} algorithm with& \textbf{0.413} & \textbf{0.468} & \textbf{0.553} & \textbf{0.390} & \textbf{0.442} & \textbf{0.531} \\
method&  disambiguation algorithm &&&&&& \\ \hline

SemCat & \SEMCATBOLD{}: no disambig. concept (pages) & $0.417$ & $0.463$ & $0.538$ & $0.393$ & $0.436$ & $0.516$ \\
method&  weighted using their rank $1/2^i$ &&&&&&  
 \\ \hline

SemCat & \SEMCATBOLD{}: no disambig. concept (pages) 	& $0.413$ & $0.464$ & $0.553$ & $0.390$ & $0.437$ & $0.531$ \\
method&   weighted using their rank $1/i$ &&&&&&  \\ \hline

SemCat &\SEMCATBOLD{}: no disambig. concept (pages) & $0.409$ & $0.442$& $0.513$ & $0.386$ & $0.416$ & $0.492$ \\
%SemCat &\SEMCATBOLD{}: no disambig. concept (pages) & $0.409$ & $0.442$*& $0.513$* & $0.386$ & $0.416$* & $0.492$* \\ -- opcja z gwiazkami od testow

method&  weighted uniformly &&&&&&  
 
 \\ \hline \hline

%&Bez dezamb.&u¿ywam 10 s³ów/10 fraz  remove false -- I.K.& $0.428$	& $0.471$	& $0.569$	& $0.394$	& $0.436$	& $0.531$  \\ \hline
%&Bez dezamb.&u¿ywam 10 s³ów/10 fraz  remove true --I.K.& $0.425$	& $0.436$	& $0.542$	& $0.38$	& $0.4$	& $0.492$  \\ \hline

% Kategoryzator&nie&Tylko s³owa/frazy jednozn.& && &&& \\
% %& $0.278$	& $0.37$	& $0.552$	& $0.253$	& $0.333$	& $0.5$\\
% && tylko dok³. dopasowania: nie&&&&&&\\
% \hline
% Kategoryzator&nie& Tylko s³owa/frazy jednozn.  & && &&& \\
% %& $0.312$	& $0.393$	& $0.536$	& $0.285$	& $0.357$	& $0.484$\\
% && tylko dok³. dopasowania: tak &&&&&&
%   \\ \hline

 %& $0.354$ & $0.49$ & $0.636$  & $0.252$	& $0.402$	& $0.578$\\ \hline

%  Ensemble of & 25x(B,W)  & $0.507$ & $0.6207$ & $0.682$ & $0.389$ & $0.517$ & $0.601$\\
%   classifiers & (kateg lvl 1) N=200 &&&&&&   \\ \hline
%  % Ensemble of & 25x(B,W)  & 0 & 0 & 0 &  $0.389706$ & $0.517345$ & $0.601562$\\
%  % (other sample)&&&&&&&   \\ \hline
%   Ensemble of & 25x(B,W)  & $0.538189$ & $0.632332$ & $0.684769$ & $0.389706$ & $0.437406$ & $0.53125$\\
%     classifiers &(kateg lvl 1)  N=70 &&&&&&   \\ \hline
%  Ensemble of & 25x(B,W) &  $0.580213$ & $0.686589$ & $0.752699$ & $0.507353$ & $0.647059$ & $0.71875$\\
%    classifiers & (kateg lvl 2) N=200 &&&&&&   \\ \hline

%% DOBRE ODKOMENTOWAC 

 Classifier&Balanced Winnow (avg of 25) L=2 S=200 & $0.547$ & \textbf{0.665} & \textbf{0.712}& $0.488$ & \textbf{0.616} & \textbf{0.669}\\    
 Classifier&Bayes (avg of 25) L=2 S=200  & $0.381$ & $0.473$ & $0.572$& $0.282$ & $0.394$ & $ 0.503$ \\   
  Classifier & LLDA  (avg of 25) L=2 S=200 & $0.437$ & $0.553$ & $0.694$& $0.385$ & $0.505$ & $ 0.652$ \\
 Classifier  & \SEMCLABOLD{} (avg of 25)& \textbf{0.558} & $0.638$ & $0.698$& \textbf{0.508} & $0.589$ & $ 0.654$ \\ \hline  \hline
 
 Ensemble &  25x(B,W) L=1 S=50 & $0.558$ & $0.652$ & $0.682$& $0.434$ & $0.558$ & $ 0.602$ \\
 Ensemble &  25x(B,W) L=1 S=100 & $0.540$ & $0.621$ & $0.667$& $0.415$ & $0.516$ & $ 0.578$ \\
 Ensemble &  25x(B,W) L=1 S=200 & $0.503$ & $0.619$ & $0.672$& $0.379$ & $0.519$ & $ 0.578$ \\
 Ensemble &  25x(B,W) L=2 S=50 & $0.577$ & $0.684$ & $0.722$& $0.515$ & $0.65$ & $ 0.68$ \\
 Ensemble &  25x(B,W) L=2 S=100 & $0.578$ & $0.698$ & $0.731$& $0.507$ & $0.655$ & $ 0.68$ \\
 Ensemble &  25x(B,W) L=2 S=200 & \textbf{0.598} & \textbf{0.699} & \textbf{0.753}& \textbf{0.518} & \textbf{0.656} & \textbf{0.711} \\ \hline
 
 Ensemble & 25x\SEMCLABOLD{} L=2 S=200 & $0.556$ & $0.637$ & $0.694$& $0.511$ & $0.59$ & $ 0.648$ \\ \hline
 Ensemble & 25x(B, W, \SEMCLABOLD{}) L=2 S=200  & \textbf{0.595} & \textbf{0.718} & \textbf{0.787}& \textbf{0.544} & \textbf{0.685} & \textbf{0.758}   \\ \hline \hline

Ensemble & 25x(B,W) L=2 S=200 + LLDA: 5.0 & $0.572$ & $0.689$ & $0.748$& $0.500$ & $0.646$ & $ 0.711$ \\
% & + LLDA: 5.0 &&&&&&   \\ 

Ensemble & 25x(B,W) L=2 S=200 + LLDA: 10.0 & $0.566$ & $0.693$ & $0.748$& $0.496$ & $0.653$ & $ 0.711$ \\
% & + LLDA: 10.0 &&&&&&   \\ 

Ensemble & 25x(B,W) L=2 S=200 + LLDA: 15.0 & $0.565$ & $0.685$ & $0.744$& $0.500$ & $0.646$ & $ 0.703$ \\
% & + LLDA: 15.0 &&&&&&   \\ 

Ensemble & 25x(B,W) L=2 S=200 + LLDA: 20.0 & $0.545$ & $0.666$ & $0.740$& $0.485$ & $0.624$ & $ 0.695$ \\ \hline
\hline
%&& Podst no disambig 15/15 & $0,387$	& $0,429$	& $0,548$	& $0,35$	& $0,393$	& $0,516$\\ %%to chyba trzeba przeliczyc jeszcze raz
%Kategoryzator&& usuwamy pozosta³e dopasowania: nie; bez s³ów kluczowych &&&&&&
%  \\ \hline

\end{tabular}

\label{tab:varius_methods_dmoz_small_part1}
\end{table*}
%\end{longtable}

%
%
%
%
%
%

\begin{table*}[h]

\centering
%\large
%\small
%\footnotesize
%\scriptsize
%|p{4.3cm} -- koluma o stalej szerokosci w cm
\caption{Average values of various precision measures for \textit{DMOZ small} dataset. Parameter $L$ stands for a level of \W{} documents used for training sample, $S$ is a sample size per each group of documents. 25x(B,W) stands for an ensemble of 25 Bayes an 25 Balanced Winnow classifiers. Vector of numbers that follows the SemCat represents weights attached to top-3 categories inserted into committee.}

\begin{tabular}{|l@{}|l@{}|c@{}|c@{}|c@{}|l@{}|l@{}|l@{}|}

\hline
\multicolumn{2}{|@{}c@{}|}{} & \multicolumn{3}{c|}{Lin precision} & \multicolumn{3}{c|}{Precision} \\
\hline
Method & Description &\footnotesize{short}&\footnotesize{medium}&\footnotesize{long} &\footnotesize{short}&\footnotesize{medium}&\footnotesize{long} \\
 \hline
     
Heterogen. & 25x(B,W) L=2 S=50+\SEMCATBOLD{}:(7,5,3) & $0.626$ & $0.715$ & $0.74$& $0.577$ & $0.715$ & $ 0.734$ \\
% & + SemCat:(7,5,3)  &&&&&&   \\ \hline

Heterogen. &    25x(B,W) L=2 S=50+\SEMCATBOLD{}:(10.5,7.5,4.5) & $0.638$ & $0.732$ & $0.759$& $0.544$ & $0.700$ & $ 0.719$ \\
% & + SemCat:(10.5,7.5,4.5) &&&&&&   \\ \hline

Heterogen. &   25x(B,W) L=2 S=50+\SEMCATBOLD{}:(14,10.6) & $0.622$ & $0.742$ & $0.771$& $0.537$ & $0.652$ & $ 0.711$ \\
% & + SemCat:(14,10.6) &&&&&&   \\ \hline

Heterogen. &   25x(B,W) L=2 S=50+\SEMCATBOLD{}:(17.5,12.5,7.5) & $0.610$ & $0.732$ & $0.770$& $0.522$ & $0.620$ & $ 0.672$ \\
% & + SemCat$:(17.5,12.5,7.5)$ &&&&&&   \\ \hline

Heterogen. &    25x(B,W) L=2 S=100+\SEMCATBOLD{}:(7,5,3)  & $0.621$ & $0.735$ & $0.784$& $0.577$ & $0.718$ & $ 0.773$ \\
% & + SemCat$:(7,5,3)$  &&&&&&   \\ 

Heterogen. &    25x(B,W) L=2 S=100+\SEMCATBOLD{}:(10.5,7.5,4.5) &$0.645$ & $0.746$ & $0.791$& \textbf{0.588} & $0.695$ & \textbf{0.766} \\
% & + SemCat:(10.5,7.5,4.5) &&&&&&   \\ 

Heterogen.&   25x(B,W) L=2 S=100+\SEMCATBOLD{}:(14,10.6) &$0.634$ & $0.752$ & $0.805$& $0.559$ & $0.671$ & $ 0.727$ \\
% & + SemCat:(14,10.6) &&&&&&   \\ 

Heterogen. &   25x(B,W) L=2 S=100+\SEMCATBOLD{}:(17.5,12.5,7.5) & $0.645$ & $0.744$ & \textbf{0.809}& $0.522$ & $0.629$ & $ 0.667$ \\
% & + SemCat:(17.5,12.5,7.5) &&&&&&   \\ 

Heterogen. &   25x(B,W) L=2 S=200+\SEMCATBOLD{}:(7,5,3) & $0.635$ & $0.731$ & $0.777$& $0.577$ & $0.722$ & $ 0.742$ \\
% & + SemCat:(7,5,3)  &&&&&&   \\ l

Heterogen. &    25x(B,W) L=2 S=200+\SEMCATBOLD{}:(10.5,7.5,4.5) & \textbf{0.645} & $0.754$ & $0.777$& $0.581$ & \textbf{0.725} & $ 0.758$ \\
% & + SemCat:(10.5,7.5,4.5) &&&&&&   \\ 

Heterogen.&   25x(B,W) L=2 S=200+\SEMCATBOLD{}:(14,10.6) & 0.644 & \textbf{0.757} & $0.780$& $0.562$ & $0.688$ & \textbf{0.765} \\
% & + SemCat:(14,10.6) &&&&&&   \\ 

Heterogen. &   25x(B,W) L=2 S=200+\SEMCATBOLD{}:(17.5,12.5,7.5)& $0.488$ & $0.570$ & $0.619$& $0.426$ & $0.508$ & $ 0.586$ \\
\hline
\hline

Heterogen. & 25x(B,W) L=2 S=200  & $0.623$ & $0.718$ & $0.776$& $0.581$ & $0.712$ & $ 0.750$ \\
& + \SEMCATBOLD{}:(10.5,7.5,4.5) + LLDA: 10.0 &&&&&&   \\ \hline

Heterogen. & 25x(B,W) L=2 S=200  & $0.591$ & $0.703$ & $0.767$& $0.555$ & $0.689$ & $ 0.758$ \\
& + \SEMCATBOLD{}:(10.5,7.5,4.5) + LLDA: 15.0 &&&&&&   \\ \hline

Heterogen. & 25x(B,W) L=2 S=200  & $0.57$ & $0.689$ & $0.751$& $0.537$ & $0.679$ & $ 0.742$ \\
& + \SEMCATBOLD{}:(10.5,7.5,4.5) + LLDA: 20.0 &&&&&&   \\ \hline

\end{tabular}

\label{tab:varius_methods_dmoz_small_part2}
\end{table*}
%\end{longtable}

\subsection{Semantic gap problem}
\label{sec:res_semantic_gap}

As visible in  Tables \ref{tab:semantic-gap-precision-CLASSICAL-METHODS}, \ref{tab:semantic-gap-precision-ENSAMBLES} in case of the semantic gap problem, semantic methods and committees lead to much
better results than traditional classifiers, even if the latter are
operating on the modified representation (bag of categories instead of
bag of words).

%all cases the \SEMCLA{} method outperforms all the other, 
%standard classification methods. 
It can be seen that the usage of terms alone gives poor results
when semantic gap occurs.
Classical methods are most helped if categories are provided for the training purposes,
but the usage of concepts is only half the way as good. 
This means actually that our \SEMCLA{} algorithm 
uses a much deeper insight into the document content than just a category label assignment. 

It is also worth to stress the fact that however \SEMCLA{} (contrary to
\SEMCAT{}) is supervised, it can also be used in unsupervised version.
For such a setting, instead of using unobservable document labels
as training classes (cf. Figure \ref{fig:semantic_classification}), one can use document clusters,
where clustering is also based on the semantic categorization (\SEMCAT{} algorithm) 
and applies semantic similarity measures defined in Section \ref{sec:semanticSimilarity}. We are going to investigate this direction more deeply in the
future, since it has a big advantage in cases where document labels
are unavailable and training set cannot be created (e.g. collections
of web pages).

\begin{table*}[h]
\caption{Average values of precision measure for classical methods: Bayes (B), Balanced Winnow (W).}
\centering
\begin{tabular}{|l||c|c c c|}
\hline
 & \multicolumn{4}{|c|}{ Classification} \\ 
 \cline{3-5} 
 & &terms &categories& concepts \\
\hline
Bankier: & Bayes& 0.397 & 0.634 &0.376 \\
 Business Biznes		& Winnow& 0.367 & 0.546&  0.323 \\

\hline		
		Forsal: & Bayes& 0.602& 0.910& 0.620 	\\
Currencies		& Winnow& 0.720 & 0.870 & 0.498  \\
\hline

 		Forsal: & Bayes& 0.847& 0.952& 0.814 \\%0.970 (centroid)	
 Finances		& Winnow&  0.832 & 0.874& 0.695 \\

 \hline
 Gynecology& Bayes& 0.404 & 0.505 & 0.233 \\%0.519 (centroid)
 		& Winnow& 0.074 & 0.205& 0.219 \\
 \hline		
 		Cardiology& Bayes& 0.782& 0.746 & 0.502 \\
 		& Winnow& 0.350  & 0.438 & 0.427 \\
 		\hline		
 		Oncology& Bayes&  0.758& 0.824& 0.526 \\
 		& Winnow& 0.227 & 0.627& 0.390 \\
 \hline
\end{tabular}

\label{tab:semantic-gap-precision-CLASSICAL-METHODS}
\end{table*}

\begin{table*}[h]
\caption{Average values of precision measure for: ``semantic classification'' (\SEMCLA), ensemble of \SEMCLA,  
 ensemble of %classifiers consisting of classical methods:
 Bayes (B), Balanced Winnow (W) and for the heterogeneous committee.
}
\centering
%\begin{tabular}{|l||@{}c@{}|c  c c||c|c|c|c|}
\begin{tabular}{|c|c|c|c|c|}

\hline
  & \SEMCLABOLD{} &25x\SEMCLABOLD{}&25x(B,W)&25x(B,W,\SEMCLABOLD{})\\

\hline

Bankier (Business Biznes):& {0.752}&0.830 &0.789&0.855\\ %0.611 (centr.)
 \hline
 		Forsal (Currencies):  &	{0.972} &0.983& 0.995  &0.999\\
 \hline		
 		Forsal (Finances): &{0.979}&0.986&0.965&0.986\\

 \hline
 Gynecology& {0.842}	&0.844 &0.732&0.833\\
 \hline		
 		Cardiology& {0.900}&$0.891$&$0.895$&$0.916$\\
 		\hline		
 		Oncology& {0.868} & 0.879&0.856&0.904\\
 \hline
\end{tabular}

\label{tab:semantic-gap-precision-ENSAMBLES}
\end{table*}

\section{Conclusions} \label{sec:conclusions}

In this paper we demonstrated the value of semantic approach to the task of
document classification. 
In particular we show here that an unsupervised approach to the classification is possible when using semantic approach. This may be considered as an interesting result by itself.
Acknowledgedly, the semantic classifier we introduce does not perform as well as ensembles of traditional classifiers but apparently an inclusion of a semantic categorizer into such an ensemble is capable of significant improvement of its performance in classic classification tasks.

Intuitively, one would imagine that a classifier  incorporating semantic information should be superior to traditional classifiers that do not use such information. As we see from our experiments it is not that obvious. Though semantic classifier proved to be a competitor for individual classic classifiers, ensembles of classic classifiers can beat it. Therefore, exploitation of advantages of semantic information requires some level of sophistication and cannot be considered as obvious.

What is still more important, the semantic classifier turns out to be superior to classical approaches to classification in case of semantic gap 
between the training data and the data for which the classifier is to be applied. 
This fact opens up really new horizons for application of machine learning methods in classification of documents in cases e.g. of mergers between various corporations where the local culture leads usually to development of specific languages different between the firms. 

This research opens up a number of further interesting areas of research.
Semantic approach (in its base, unsupervised setting) 
could be tested also for clustering tasks under semantic gap scenario as well as to mixtures of classification and clustering.  

\bibliography{bibliografia}

\begin{thebibliography}{10}

\bibitem{SIIS:2011}
Ciesielski, K., Borkowski, P., Klopotek, M.A., Trojanowski, K., Wysocki, K.:
\newblock Wikipedia-based document categorization.
\newblock In: Security and Intelligent Information Systems, SIIS 2011, Warsaw,
  Poland, June 13-14, 2011. (2011)  265--278

\bibitem{Sebastiani:2002}
Sebastiani, F.:
\newblock Machine learning in automated text categorization.
\newblock ACM Comput. Surv. \textbf{34}(1) (March 2002)  1--47

\bibitem{Sebastiani05textcategorization}
Sebastiani, F.:
\newblock Text categorization.
\newblock In: Text Mining and its Applications to Intelligence, CRM and
  Knowledge Management, WIT Press (2005)  109--129

\bibitem{SeppanenBM03}
Sepp{\"a}nen, J.K., Bingham, E., Mannila, H.:
\newblock A simple algorithm for topic identification in 0-1 data.
\newblock In Lavrac, N., Gamberger, D., Blockeel, H., Todorovski, L., eds.:
  PKDD. Volume 2838 of Lecture Notes in Computer Science., Springer (2003)
  423--434

\bibitem{Medelyan08}
Medelyan, O., Witten, I.H., Milne, D.:
\newblock Topic indexing with wikipedia.
\newblock In: Proceedings of the first AAAI Workshop on Wikipedia and
  Artificial Intelligence (WIKIAI'08). (2008)

\bibitem{Milne08}
Milne, D., Witten, I.H.:
\newblock An effective, low-cost measure of semantic relatedness obtained from
  wikipedia links.
\newblock In: Proceedings of the first AAAI Workshop on Wikipedia and
  Artificial Intelligence (WIKIAI'08). (2008)

\bibitem{MilneW08}
Milne, D.N., Witten, I.H.:
\newblock Learning to link with wikipedia.
\newblock In: Proceedings of the 17th {ACM} Conference on Information and
  Knowledge Management, {CIKM} 2008, Napa Valley, CA, {USA}, October 26-30,
  2008, ACM (2008)  509--518

\bibitem{MihalceaC07}
Mihalcea, R., Csomai, A.:
\newblock Wikify!: linking documents to encyclopedic knowledge.
\newblock In: Proceedings of the Sixteenth {ACM} Conference on Information and
  Knowledge Management, {CIKM} 2007, Lisbon, Portugal, November 6-10, 2007, ACM
  (2007)  233--242

\bibitem{sydow:debora-ismis12}
Wroblewska, A., Sydow, M.:
\newblock Debora: Dependency-based method for extracting entity-relationship
  triples from open-domain texts in polish.
\newblock In Chen, L., Felfernig, A., Liu, J., Ras, Z., eds.: Foundations of
  Intelligent Systems. Volume 7661 of Lecture Notes in Computer Science.
\newblock Springer Berlin Heidelberg (2012)  155--161

\bibitem{ramakrishna2012survey}
Ramakrishna~Murty, M., Murthy, J., Prasad~Reddy, P., Satapathy, S.:
\newblock A survey of cross-domain text categorization techniques.
\newblock In: Recent Advances in Information Technology (RAIT), 2012 1st
  International Conference on, IEEE (2012)  499--504

\bibitem{wang2008using}
Wang, P., Domeniconi, C., Hu, J.:
\newblock Using wikipedia for co-clustering based cross-domain text
  classification.
\newblock In: Data Mining, 2008. ICDM'08. Eighth IEEE International Conference
  on, IEEE (2008)  1085--1090

\bibitem{nguyen2010bridging}
Nguyen, C.T.:
\newblock Bridging semantic gaps in information retrieval: Context-based
  approaches.
\newblock ACM VLDB \textbf{10} (2010)

\bibitem{ContentCategorizationRafi2012}
Rafi, M., Hassan, S., Shaikh, M.S.:
\newblock Content-based text categorization using wikitology.
\newblock CoRR \textbf{abs/1208.3623} (2012)

\bibitem{SemanticSimilarityMetric}
Pirr{\`o}, G., Seco, N.:
\newblock Design, implementation and evaluation of a new semantic similarity
  metric combining features and intrinsic information content.
\newblock In: On the Move to Meaningful Internet Systems. Volume 5332 of LNCS.,
  Springer (2008)  1271--1288

\bibitem{Aas99textcategorisation}
Aas, K., Eikvil, L.:
\newblock Text categorisation: A survey.
\newblock Report No. 941 (June 1999) ISBN 82-539-0425-8.

\bibitem{GroveBWinnow:2001}
Grove, A.J., Littlestone, N., Schuurmans, D.:
\newblock General convergence results for linear discriminant updates.
\newblock Mach. Learn. \textbf{43}(3) (June 2001)  173--210

\bibitem{LittlestoneBWinnow1988}
Littlestone, N.:
\newblock Learning quickly when irrelevant attributes abound: A new
  linear-threshold algorithm.
\newblock Machine Learning \textbf{2} (1988)  285--318

\bibitem{Rosenblatt:1957}
Rosenblatt, F.:
\newblock The perceptron: {A} perceiving and recognizing automaton.
\newblock Technical Report 85-460-1, Ithaca, New York (January 1957)

\bibitem{Blei:2003}
Blei, D.M., Ng, A.Y., Jordan, M.I.:
\newblock Latent dirichlet allocation.
\newblock J. Mach. Learn. Res. \textbf{3} (March 2003)  993--1022

\bibitem{steyvers2006probabilistic}
Steyvers, M., Griffiths, T.:
\newblock Probabilistic topic models.
\newblock In Landauer, T., Mcnamara, D., Dennis, S., Kintsch, W., eds.: Latent
  Semantic Analysis: A Road to Meaning.
\newblock Laurence Erlbaum (2006)

\bibitem{Ramage:2009}
Ramage, D., Hall, D., Nallapati, R., Manning, C.D.:
\newblock Labeled lda: a supervised topic model for credit attribution in
  multi-labeled corpora.
\newblock In: Proceedings of the 2009 Conference on Empirical Methods in
  Natural Language Processing: Volume 1 - Volume 1. EMNLP '09, Stroudsburg, PA,
  USA, Association for Computational Linguistics (2009)  248--256

\bibitem{blei2007supervised}
Blei, D.M., McAuliffe, J.D.:
\newblock Supervised topic models.
\newblock In: NIPS. (2007)

\bibitem{Ramage:2011}
Ramage, D., Manning, C.D., Dumais, S.:
\newblock Partially labeled topic models for interpretable text mining.
\newblock In: Proceedings of the 17th ACM SIGKDD international conference on
  Knowledge discovery and data mining. KDD '11, New York, NY, USA, ACM (2011)
  457--465

\bibitem{Elements:Hastie}
Hastie, T., Tibshirani, R., Friedman, J.H.:
\newblock {The Elements of Statistical Learning}.
\newblock Springer (2008)

\end{thebibliography}
\bibliographystyle{splncs}

\end{document}